\documentclass[
twocolumn,
 amsmath,amssymb,
 aps,
pra,
]{revtex4-2}

\usepackage{color,graphicx}
\usepackage[colorlinks=true,linkcolor=blue,citecolor=blue,urlcolor=blue]{hyperref}
\usepackage[capitalise]{cleveref} 
\usepackage{siunitx}
\usepackage{mathrsfs}

\DeclareMathOperator{\STr}{STr} 
\DeclareMathOperator{\tr}{tr} 
\DeclareMathOperator{\Tr}{Tr} 
\DeclareMathOperator{\re}{[1]} 
\DeclareMathOperator{\im}{[2]} 
\DeclareMathOperator{\id}{\mathrm{I}}
\DeclareMathOperator{\idj}{\mathrm{J}}

\newcommand{\vol}{\mathcal{V}}
\newcommand{\W}{W_k}
\newcommand{\V}{U_k}
\newcommand{\x}{{\mathbf x}}
\newcommand{\p}{{\mathbf p}}
\newcommand{\q}{{\mathbf q}} 
\newcommand{\bpsi}{\bar{\psi}}
\newcommand{\Z}{{Z_k}}
\newcommand{\A}{{Y_k}}
\newcommand{\E}{a}

\begin{document}

\title{Fluctuation-induced first-order superfluid transition in unitary $\mathrm{SU}(N)$ Fermi gases}

\author{Georgii Kalagov}
\email{kalagov@theor.jinr.ru}
\homepage{https://orcid.org/0000-0003-2208-0040}
\affiliation{Joint Institute for Nuclear Research, Joliot-Curie, 6, 141980 Dubna, Russia
}

\begin{abstract}

We investigate the superfluid phase transition in an $\mathrm{SU}(N)$-symmetric Fermi gas with $N$ distinct spin states using the functional renormalization group. To capture pairing phenomena beyond mean-field theory, we introduce an auxiliary bosonic field and employ  the leading order of the derivative expansion of the partially bosonized effective average action. By discretizing the effective potential on a grid and numerically integrating the flow equations, we resolve the thermodynamic behavior near the transition. Our results reveal a fluctuation-induced first-order phase transition for $N \geq 4$, which is absent at the mean-field level. In the unitary regime, we provide quantitative predictions for the critical temperature, as well as for the discontinuities in the superfluid gap and entropy density as functions of $N$. With increasing $N$, the critical temperature decreases, while the discontinuities become more pronounced, indicating a stronger first-order transition.   
\end{abstract}

\maketitle

\section{Introduction}

Interacting Fermi systems lie at the heart of phenomena as disparate as neutron‑star matter and high‑temperature superconductivity. Ultracold atomic gases now offer a pristine, fully tunable platform for probing  such many‑body physics with unprecedented control. Whereas most experimental work has focused on two‑component (spin-$1/2$) mixtures, the recent realization of gases with $\mathrm{SU}(N)$ symmetry opens a route to exploring enlarged spin manifolds where quantum statistics and symmetry intertwine in qualitatively new ways \cite{Wu2010,Cazalilla2014,Gorshkov2014,Mukherjee2025}. Alkaline-earth-like atoms are ideal for this purpose: their closed electronic shells quench the total electronic angular momentum,  leaving the atomic spin purely nuclear. Consequently, an atom with nuclear spin $I$ possesses $N = 2I + 1$ degenerate hyperfine states, implementing an $\mathrm{SU}(N)$‑symmetric Fermi system in the laboratory. Prominent examples already accessible experimentally include $^{87}\mathrm{Sr}$ ($I = 9/2$) and $^{173}\mathrm{Yb}$ ($I = 5/2$) \cite{Desalvo2010,Taie2012,Taie2010,Krauser2012,Pagano2014,Zhang2014}. Systems with enlarged $\mathrm{SU}(N)$ symmetry exhibit a rich variety of many-body phenomena. Foremost  among these is the superfluid phase transition,  which has been the focus of extensive theoretical and experimental investigation
\cite{Ho1999,Honerkamp2004,Cherng2007,Capponi2008,He2006,Rapp2007,Ozawa2010,Schlottmann2014,Zhang2021, Perlin2022}. 

Previous field-theoretical studies of the superfluid transition in multicomponent systems have employed a Landau–Ginzburg–type effective model involving an antisymmetric tensor order parameter \cite{Ho1999, komarova2013}. A standard renormalization group analysis, based on multi-loop calculations within the $\varepsilon$-expansion, revealed the absence of infrared (IR) stable fixed points for $N \geq 4$ \cite{kalagov2016,Nalimov2020,bednyakov2021}. As a result, the system does not exhibit critical scaling and instead shows non-universal behavior qualitatively consistent with a first-order phase transition. In the two-component case ($N=2$), by contrast, the system undergoes a continuous phase transition characterized by the standard set of critical exponents associated with the  $XY$ model universality class. Key thermodynamic properties of the ordered phase---such as the critical temperature, the energy gap, and related observables---depend on the interparticle interaction, encoded in the $s$-wave scattering length $a_s$. In the unitary regime, where  $1/a_s = 0$, these quantities have been extensively investigated in experiments \cite{Ku2012}, through numerical simulations \cite{Burovski2006,Bulgac2008}, and via theoretical approaches \cite{Haussmann2007}, including those based on the functional renormalization group (FRG) method \cite{Diehl2010,Boettcher2012,Boettcher14,Fejos2014}. 
The FRG approach, being inherently nonperturbative, enables the simultaneous treatment of long-wavelength critical fluctuations and  the computation of universal exponents, as well as the analysis  of non-universal thermodynamic properties of the system within a unified framework that remains valid across both symmetric and symmetry-broken phases.

In this work, we investigate the nature of the phase transition in $\mathrm{SU}(N)$-symmetric systems of fermions with attractive interactions, going beyond mean-field theory by means of the FRG, with a focus on the unitary regime. Our aim is to demonstrate the emergence of a fluctuation-induced first-order phase transition for $N \geq 4$ components, determine the critical temperature, and track the development of discontinuities in the superfluid gap and entropy density. To this end, we solve the FRG flow equation using a partially bosonized ansatz for the effective average action, integrating out both fermionic and bosonic modes at leading order in the derivative expansion. We treat the resulting flow equation as a partial differential equation, discretize it on a grid, and obtain a global solution that reveals the onset of the first-order transition.

The paper is organized as follows. In Section~\ref{sec:model}, we introduce the microscopic fermionic model. Section~\ref{sec:mf} presents the mean-field analysis of the phase transition and the associated symmetry-breaking pattern. In Section~\ref{sec:frg}, we outline the FRG framework, specify our nonperturbative truncation of the effective average action, and derive the flow equations for the scale-dependent couplings. Section~\ref{sec:results} reports the global numerical solution of these flow equations on a  grid and presents  quantitative results for the critical temperature, as well as the discontinuities in the superfluid gap and the specific entropy. Section~\ref{sec:consl} concludes the paper and offers an outlook. In the Appendices, we provide detailed derivations of the FRG flow equations and describe the numerical methods employed in the main text.

\section{\label{sec:model} Model}

The microscopic model of an equilibrium system of $N$-component fermions (where $N$ is even) in contact with a thermal bath at temperature $T$ and chemical potential $\mu$ is defined by the local action 
\begin{equation}
\label{eq:fermi_action}
S[\psi] =\int_{X}\left[ \bpsi_a  \left(\partial_0 -\nabla^2 - \mu \right)\psi_a + \frac{\lambda_0}{2}
\left(\bpsi_a \psi_a\right)^2\right].
\end{equation} 
The Grassmann fields $\psi_{a}=\psi_{a}(x_0,\x)$ are functions of the Euclidean time variable $x_0 \in [0,\beta=T^{-1}]$ and spatial coordinates $\x\in \mathbb{R}^d$, the subscript $a=1,\dots,N$ labels the internal degrees of freedom. Fermionic fields satisfy anti-periodic boundary conditions $\psi_{a}(0,\x)=-\psi_{a}(\beta,\x)$, and  $\bpsi_a$ stands for the complex conjugate counterpart. The negative magnitude $\lambda_0$ of the bare zero-range pairwise interaction requires an appropriate renormalization. Herein, one uses the abbreviation $X \equiv (x_0, \x)$, $\int_{X} \equiv \int_{0}^{\beta} dx_0 \int d^d \mathrm{x}$, and $\partial_0 \equiv \partial/\partial x_0$.  We leave the number of dimensions $d$ unspecified, focusing on $d=3$ later. We adopt a system of units where $k_B = \hbar = 2 m = 1$, with $m$ denoting the particle mass.

We consider a homogeneous system in the grand canonical ensemble. The thermodynamic  properties are thus obtained from the grand potential  $\Omega$, defined by 
\begin{equation}
   \exp(-\beta \Omega) = \mathcal{N} \int \mathcal{D}\psi \mathcal{D}\psi^{\dag} \exp(-S[\psi]),
\end{equation}
 where the normalization factor $\mathcal{N}$ is chosen so that $\Omega=0$ for the noninteracting system. In other words, the ideal Fermi gas contribution has been subtracted from  $\Omega$.  

As a standard procedure, the four-fermion interaction is decoupled in the particle-particle channel by introducing an auxiliary bosonic field $ \phi =  \phi_{a b}(X)$, which represents the paired state. The bosonic mode $\phi$ is a complex antisymmetric field of size $N \times N$, subject to the periodic boundary condition $\phi(0,\x)= \phi(\beta,\x)$. The Hubbard–Stratonovich transformation \cite{komarova2013} results in the following microscopic action 
\begin{align}
\label{eq:fermi_bose_action}
\notag
     S[\psi, \phi] = &\int_{X}  \Big[  \bpsi_a  \left(\partial_{0} -\nabla^2 - \mu \right) \psi_a + \frac{m_0^2}{2} \tr( \phi^{\dag} \phi)  \\
     &+ \frac{g_0}{2} (  \bpsi_a \phi_{a b} \bpsi_b  +  \psi_a \phi^{\dag}_{a b} \psi_b) \Big]. 
\end{align}
Performing Gaussian integration over the $ \phi$-fields, we restore the original action \cref{eq:fermi_action}, with the coupling constant $\lambda_0 = - g_0^2/m_0^2 < 0$.  Integrating out the $\psi$-fields yields a  bosonic path integral representation for the grand potential 
\begin{equation}
   \exp(-\beta \Omega) =  \int \mathcal{D} \phi \mathcal{D} \phi^{\dag} \exp(-S[\phi]).
\end{equation}   
The resulting nonlocal action can be written as
\begin{align}
\label{eq:bose_action}
    S[\phi] = \frac{m_0^2}{2} \int_{X} \tr( \phi^{\dag} \phi) -\frac{1}{2} \Tr \ln (\hat{1} - \hat{G}_{\psi}  \hat{K}_{ \phi}).
\end{align}
Here, the matrix elements of the free propagator  $\hat{G}_{\psi}$  in $2N$-dimensional space of vectors $\Psi = (\psi_1,\dots,\psi_N,  \bpsi_1,\dots,\bpsi_N)^T$ are given by
\begin{align}
   G_{\psi}(Q, Q') = \delta(Q-Q') \begin{pmatrix}
0 &  [i q_0 - \xi_{\q}]^{-1} \\
 [i q_0 + \xi_{\q}]^{-1} & 0 \\
\end{pmatrix}\otimes \id_N,
\end{align}
where $\id_N$ is the unit matrix of size $N\times N$, $\xi_{\q} = \q^2-\mu$ is the energy of the free single particle states measured from the fermion chemical potential, $q_0 \in \pi T (2\mathbb{Z}+1)$ are fermionic Matsubara frequencies, and $ \delta(Q-Q')= \beta (2\pi)^d  \delta_{q_0 q_0'} \delta(\q-\q')$.   In the position representation, the matrix element of $\hat{K}_{ \phi}$ is simply
\begin{align}
  K_{ \phi}(X, X') = \delta(X-X') \begin{pmatrix}
 \phi^{\dag}(X) & 0 \\
 0 & \phi(X) \\
\end{pmatrix}. 
\end{align}

Note that no kinetic term for the collective $\phi$-field appears in the classical action \cref{eq:fermi_bose_action}. This term arises dynamically when fermion fluctuations are integrated out and can be derived by expanding the action \cref{eq:bose_action} in terms of the $\phi$-field and its derivatives: $\partial_0 \phi$ and $\nabla \phi$. The Schwinger equations \cite{Vasiliev} demonstrate that a finite expectation value of this field below the critical temperature corresponds to the anomalous composite operator $\langle \phi_{a b} \rangle \sim \langle \psi_a \psi_b \rangle$. It can be identified as the order parameter of the phase transition, which is associated with $\mathrm{U}(N)$ symmetry breaking and  results in a finite gap in the fermionic spectrum.

\section{\label{sec:mf} Mean field and symmetry breaking}

The underlying action \cref{eq:fermi_action} possesses a global unitary symmetry: it remains invariant under the transformation  $\psi \to   h \psi$,  for $h \in \mathrm{U}(N)$. The composite field $ \phi$ transforms as a rank-2 antisymmetric tensor $ \phi \to  h  \phi  h^{T}$. For convenience, we will parameterize the classical action \cref{eq:bose_action} using the ``gap field'', which is achieved by the formal rescaling $\phi \to \phi/g_0$.  Let $ \phi_0$ be a homogeneous mean-field vacuum expectation value that is given by the global minimum of \cref{eq:bose_action}. There is a unitary rotation that brings the complex antisymmetric matrix into block-diagonal form  $ \phi_0= \mathrm{diag}\{ \Delta_{0,1} \epsilon,\dots, \Delta_{0, N/2} \epsilon\}$, where $\epsilon$ is the antisymmetric unit tensor of size $2\times 2$, normalized as $\epsilon_{12}=1$, and the $\Delta_0$-amplitudes are non-negative real numbers \cite{Zumino1962}. Then, the specific classical potential $V_{\text{cl}}(\phi_0) = S[\phi_0]/\vol_{\beta}$, where the total volume $\vol_{\beta} = \beta \vol=\beta \int d^d x$, determines the mean-field value of the grand thermodynamical potential: $\Omega = \vol \min_{ \phi_0}V_{\text{cl}}(\phi_0)$. The classical potential on the considered background can be written as  
\begin{align}
   V_{\text{cl}}(\phi_0)  = - \sum_{i=1}^{N/2} \left[ \frac{\Delta_{0,i}^2}{ \lambda_0}  +   \int_Q \ln \left( 1+ \frac{\Delta_{0,i}^2}{q_0^2+\xi_\q^2} \right) \right],
\end{align}
where 
\begin{align}
    \int_{Q}  =  T \sum_{q_0 }\int_\q =  T \sum_{q_0 }\int\frac{d^dq}{(2\pi)^d}, \quad Q  = (q_0, \q).
\end{align}
Here, the  renormalized counterpart of $\lambda_0$  can be defined by the low energy limit of the two-body problem in vacuum \cite{abrikosov1975}
\begin{align}
\label{eq:l_renorm}
    \frac{1}{\lambda} = \frac{1}{\lambda_0} + \int_{|\q|<\Lambda}\frac{1}{2  \varepsilon_{\q}}, \quad \varepsilon_{\q} = \q^2,
\end{align}
where the momentum integral is regularized for $d \geq 2$ by an ultraviolet (UV) cutoff $\Lambda$, which is related to the inverse of the range of the interaction potential. Removing the bare interaction from the problem ensures that the physical quantities depend solely on a single amplitude $\lambda$. For $d=3$, this amplitude is directly related to the $s$-wave scattering length $a_s$ and given by $\lambda = 8 \pi a_s  <0$. Since $\lambda$ is kept constant as the cutoff is taken to infinity, the magnitude  $\lambda_0 \to 0^{-}$.

Moving on, we proceed to minimize the potential: $ \partial V_{\text{cl}} / \partial \Delta_{0,i} = 0$, for $i = 1,\dots,N/2$, which leads to the system
\begin{align}
    \Delta_{0,i} \left[ \frac{1}{ \lambda_0} +  \int_Q \frac{1}{\left( q_0^2+\xi_\q^2+ \Delta_{0,i}^2 \right)}  \right] = 0.
\end{align}
 There are several  ways to satisfy these equations. For high temperatures $T > T_c$ the expression in the square brackets is nonzero, resulting in only the trivial solution $\Delta_{0,i} = 0$ for $\forall i$. For low temperatures $T < T_c$  there exist $k$ nonvanishing solutions $\Delta_{0,i} \ne 0$, while the remaining $N/2 - k$ magnitudes vanish $\Delta_{0,i} = 0$. In this case, the unbroken symmetry is $\mathrm{SU}(N-2k)\otimes \mathrm{USp}(2k)$, corresponding to $k$ gapped branches in the quasi-particle spectrum and $N/2-k$ gapless states. The value of mean field critical temperature $T_c$ is determined from the standard equation
\begin{equation}
\label{eq:t_c}
    \frac{1}{ \lambda_0} +  \left.\int_Q \frac{1}{\left( q_0^2+\xi_\q^2 \right)}\right|_{T=T_c}=0,
\end{equation}
containing two divergent terms:  the infinite factor $1/\lambda_0$, \cref{eq:l_renorm}, and the divergent integral, which compensate each other. Since all non-zero $\Delta_0$-amplitudes satisfy the same equations, they are equal; therefore, the subscript ``$i$'' will be omitted below. Let us compute the minimum of the potential and determine the matrix structure of the condensate, i.e. the value of $k$, at which it attains the global minimum. For fixed $k$ it can be written as 
\begin{align}
\notag
    \min_{ \phi_0}V_{\text{cl}}(\phi_0) = - k \int_Q L \left( \frac{\Delta_0^2}{q_0^2+\xi_\q^2}  \right),
\end{align}
where the  function $L(x) = -x/(1+x) + \ln(1+x)$ is positive definite for all $x \in (0, \infty)$; therefore, the classical potential attains its global minimum at the maximum value of $k$, which is  $k = N/2$.  Thus, we have established that the vacuum expectation value is $ \phi_0= \Delta_0 \id_{N/2} \otimes \epsilon$.  In this work, for computational convenience, we adopt the symplectic basis, where $ \phi_0 = \Delta_0 \idj$ with 
\begin{align}
\idj  \equiv \epsilon\otimes \id_{N/2} = \begin{pmatrix}
 0 & \id_{N/2} \\
 -\id_{N/2} &  0  \\
\end{pmatrix}, 
\end{align}
and the unbroken subgroup is thus  $\mathrm{USp}(N) = \{ h\in \mathrm{SU}(N) \  | \ h \idj h^{T} = \idj\}$. 

This analysis shows that within naive mean-field theory fermionic systems with $\mathrm{U}(N)$-symmetric interaction undergo a continuous superfluid phase transition for any spin multiplicity $N$, and the resulting critical temperature, obtained from \cref{eq:t_c},  does not depend on $N$.

\section{\label{sec:frg}Renormalization group treatment}

This study employs the functional renormalization group to properly account for bosonic fluctuations neglected in the mean-field analysis. The essence of this approach lies in the scale-dependent effective average action $\Gamma_k$, which comprises all the fluctuations of the field model beyond the momentum scale $k$ \cite{dupuis2021, berges2002, pawlowski2007}. That is, $\Gamma_k$ is IR regularized by the cut-off function $R_k$ added to the original action $S$ in the form of a momentum-dependent mass term. The functional $\Gamma_k$ smoothly interpolates between the original microscopic action at the UV scale $k = \Lambda$, where $\Gamma_{\Lambda} = S$, 
and the quantum effective action in the IR limit $\Gamma_{k\to 0} = \Gamma$. Successive infinitesimal scale steps correspond to a continuous flow in a functional space governed by the exact equation
\begin{equation}
\label{eq:weq}
     \partial_t \Gamma_k = \frac{1}{2}  \hat{\partial}_t \STr{ \ln(\Gamma_k^{(2)}+R_k)}.
\end{equation}
Here, $t = \ln(k/\Lambda)$, and the derivative $\hat{\partial}_t$ acts exclusively on the regulator function $R_k$. The supertrace $\STr$ extends over all fermionic and bosonic degrees of freedom, while the operator $\Gamma_k^{(2)}$  denotes the second functional derivative of the effective action $\Gamma_k$ with respect to the fields of the model. The flow \cref{eq:weq} is a complex functional integro-differential equation that, while unsolvable exactly, provides a practical foundation for nonperturbative approximation schemes.

The following analysis is based  on the truncated effective average action $\Gamma_k$, which preserves all symmetries of the classical action. In this work, we employ the leading order of the derivative expansion, given by
\begin{align}
\notag
     \Gamma_k =& \int_{X}  \Big[\bpsi_a  \left(\partial_{0} -\nabla^2 - \mu \right)\psi_a \\[1ex]    \notag 
     &+\A   \tr(\phi^{\dag} \partial_{0} \phi)  - \Z \tr(\phi^{\dag} \nabla^2 \phi) + V_k(\phi^{\dag}\phi)\\[1ex]  
     & +  \frac{g_k}{2} (  \bpsi_a \phi_{a b} \bpsi_b  +  \psi_a \phi^{\dag}_{a b} \psi_b)\Big]. 
\label{eq:lpa}
\end{align}
The coefficients in the boson kinetic term are evaluated in an equilibrium background, which corresponds to the minimum of the running potential $V_k$. Since we directly associate the  expectation value of the bosonic field $\phi$ with the gap in the fermion spectrum, it is normalized in such a way that the ``Yukawa coupling'' $g_k$ is dimensionless. The key point of our analysis is to retain the fermion propagator in its microscopic form and fix the coupling $g_k$ to unity at all scales. 
Following \cite{Boettcher14}, we anticipate that the running of $g_k$ is subleading; accordingly, allowing only $\A, \Z$, and $V_k$  to evolve with $k$, as this constitutes the minimal set required for a dynamic treatment of bosonic fluctuations. To match the value of the functional $  \Gamma_{\Lambda}$ with the microscopic action \cref{eq:fermi_bose_action} at the UV scale, the initial conditions must be chosen in the form 
\begin{equation}
\label{eq:initial_cond}
    Z_{\Lambda} = 0,  \quad  Y_{\Lambda} = 0,  \quad   V_{\Lambda}(\phi^{\dag}\phi) = \frac{m_0^2}{2}  \tr(\phi^{\dag}\phi). 
\end{equation}
In our numerical implementation, the values of the couplings  $\Z$ and $\A$ are chosen to be nonzero yet sufficiently small at the outset of the renormalization group flow, where $k \sim \Lambda$. The bar parameter $m_0^2$ has to be tuned to achieve the correct vacuum scattering length as $k \to 0$.

In computing the flow equation, there is considerable freedom in the choice of the regulator function $R_k(Q)$. In the bosonic sector, the cutoff we employ is the optimized Litim regulator \cite{litim2000, litim2001}
\begin{equation}
\label{eq:litimb}
    R^{(\mathcal{B})}_k(Q) = \Z (k^2 - \q^2)\theta(k^2 - \q^2).
\end{equation}
It provides an additional mass term to the single-particle energy of states with small momenta ($q < k$), thereby suppressing them while leaving large-momentum modes ($q > k$) unaffected. For the fermionic field we choose the similar cutoff
\begin{equation}
\label{eq:litimf}
    R^{(\mathcal{F})}_k(Q) = \mathrm{sgn}(\xi_{\q}) (k^2 - |\xi_{\q}|)\theta(k^2 -|\xi_{\q}|),
\end{equation}
which regularizes around the Fermi surface \cite{Diehl2010}. In each regulator, the Heaviside step function is treated as a weak limit of smooth functions and is further specified by the condition $\theta(0)=1/2$. This allows for a semi-analytical treatment of the loop integrals in the subsequent analysis. Within this scheme, the condition required to tune $m_0^2$ is  $m_0^2/g_0^2 = 1/8 \pi a_s + \Lambda/6 \pi^2$ \cite{Boettcher14, Braaten2008}. Note that coefficient of $\Lambda$ differs from that obtained in the simple cutoff regularization \cref{eq:l_renorm}, where it is equal to $1/4 \pi^2$. In the present calculations, we restrict ourselves to the unitary limit by setting $1/a_s \to 0^{-}$. However, the equations derived below remain applicable to systems away from unitarity as well.

 \subsection{Flow of the effective potential}
 
The evolution equations for the scale-dependent couplings in \cref{eq:lpa} consist of two distinct contributions: a bosonic part ($\mathcal{B}$) and a fermionic part ($\mathcal{F}$), both arising from the supertrace in \cref{eq:weq}. We evaluate these contributions separately and express the flow of a generic coupling---denoted here by $A_k$---as $\partial_t A_k  \equiv \dot{A_k}^{(\mathcal{B})} + \dot{A_k}^{(\mathcal{F})}$. 

The flow equation for the scalar potential $V_k$ is obtained by evaluating \cref{eq:weq} for a real constant background $\phi(X) = \Phi = \mathrm{const}$ whose tensor structure corresponds to the symmetry breaking $\mathrm{U}(N) \to \mathrm{USp}(N)$.  Specifically, we take the bosonic background to be $\Phi = \Delta \idj$, where $\Delta$ is a positive magnitude, while the background value of the fermionic field is zero. The potential $V_k$ is a general function of $\mathrm{U}(N)$-invariant field monomials. Owing to the property of the background field $\Phi^{\dag} \Phi \sim \id_N$, it is convenient to choose invariants in the following manner: the quadratic invariant is defined as $\rho_1 = \tr(\phi^{\dag}\phi)$, and higher-order invariants are constructed from the powers of the traceless tensor $\phi^{\dag}\phi -  \id_N \rho_1/N$ and take the form $\rho_i = \tr(\phi^{\dag}\phi -  \id_N \rho_1/N)^i$. A useful property of such a choice is that invariants $\rho_{i \geq 2}$, when evaluated on the background field $\Phi$, are identically zero. Thus,  for an arbitrary field $\phi$, we parametrize the potential as  $V_k = V_k(\rho_1, \rho_2, \rho_3, \dots)$, where dots represent  higher-order invariants.

The projection of the equation \cref{eq:weq} onto the background $\Phi$ yields $\partial_t \V = \vol_{\beta}^{-1} \partial_t \Gamma_k|_{\phi = \Phi, \Psi = 0}$ with $\V(\rho_1) \equiv  V_k(\rho_1, 0,0, \dots)$. For the calculation of the bosonic contribution $\dot{\V}^{(\mathcal{B})}$, we adopt a convenient parametrization of the fluctuating field $\phi$ based on the symmetry breaking pattern. The dimension of the unbroken subgroup $\mathrm{USp}(N) $ is equal to $N(N+1)/2$, while the dimension of $\mathrm{U}(N)$ equals $N^2$.  We denote the unbroken generators by $\Xi_{i}$, with $i = 1, \dots,  N(N+1)/2$. The remaining ones form the set $\{\id_{N}, \Pi_{a}\}$, where the unit matrix $\id_{N}$ corresponds to $U(1)$ symmetry of global phase rotation, and $ \Pi_{a}$ are traceless hermitian generators, $a = 1,\dots,  N(N-1)/2-1$. The subgroup generators $\Xi_{i}$ satisfy the relation $ (\idj \Xi_i)^T = \idj \Xi_i$, while the broken generators, which span the Goldstone manifold $\mathrm{U}(N)/\mathrm{USp}(N)$, obey $(\idj \Pi_a)^T = -\idj \Pi_a$, and are normalized according to $\tr (\Pi_a \Pi_b) = \delta_{a b}/2$.  Using this set of generators, we parameterize the field as
\begin{align}
\label{eq:nonlin}
\notag
     \phi &=  \exp\left(\frac{i \alpha}{ \sqrt{2 N } \,\Delta}\right)  \exp\left(\frac{i \pi_a \Pi_a}{2 \Delta}\right) \Big( \Phi +\frac{\sigma}{\sqrt{2 N }} \idj  \big. \\[1ex] \notag
         &\big. + \zeta_a \Pi_a\idj \Big) \exp\left(\frac{i \pi_a \Pi^T_a}{2 \Delta}\right)
     \\[1ex]  & \approx
       \Phi + \frac{\sigma + i \alpha}{\sqrt{2 N }} \idj + (\xi_a  + i \pi_a) \Pi_a \idj.
\end{align}
Here $\sigma$, $\alpha$, $\vec{\zeta}$ and  $\vec{\pi}$ are real-valued functions of $X$. The total number of real components in \cref{eq:nonlin} is respectively $n_{\sigma} + n_{\alpha} +n_{\zeta} + n_{\pi} \equiv 1 + 1 + [ N(N-1)/2 - 1] +  [N(N-1)/2 - 1] =  N(N-1)$, which equals the number of components of a complex antisymmetric matrix. The singlet $\alpha$ and the fields $\vec{\pi}$ are the ``Goldstone modes'', while $\sigma$ and $\vec{\zeta}$ represent the ``radial modes''. We chose the normalization factors in the denominators to obtain a canonical kinetic terms for all modes, e.g. $\Z (\partial_i \alpha)^2/2$.

The decomposition \cref{eq:nonlin} allows for the diagonalization of the bosonic part $[\Gamma_k^{(2)}]^{(\mathcal{B})}$of $\Gamma_k^{(2)}$ in \cref{eq:weq}, with its representation in block-diagonal form
\begin{equation}
\label{eq:unperturb_hess}
 [\Gamma_k^{(2)}(Q,Q')]^{(\mathcal{B})} =\delta(Q+Q')  \begin{pmatrix}
 \bar{\Gamma}^{(S)}_k(Q)& 0 \\
 0 & \bar{\Gamma}^{(V)}_k(Q) \\
\end{pmatrix},
\end{equation}
with the elements 
\begin{subequations}
\label{eq:bosongama2}
\begin{align}
&\bar{\Gamma}^{(S)}_k(Q) =    \begin{pmatrix}
  \Z \q^2 +m_{\alpha}^2 & -\A  q_0 \\
 \A  q_0 &  \Z \q^2 + m_{\sigma}^2  \\
\end{pmatrix}, \\[2ex]
& \bar{\Gamma}^{(V)}_k(Q)  =    \begin{pmatrix}
  \Z \q^2  + m_{\pi}^2  & -\A  q_0 \\
 \A  q_0 &  \Z \q^2 + m_{\zeta}^2\\
\end{pmatrix} \otimes \id_{n_{\zeta}}, 
\end{align}
\end{subequations}
where  $q_0 \in 2 \pi T  \mathbb{Z}$. The eigenvalues of the ``mass matrix'' in the background $\Phi$ are given by 
\begin{subequations}
\begin{align}
&m_{\alpha}^2=m_{\pi}^2 =   \V',\\
& m_{\sigma}^2 =  \V' + 2\rho_1  \V'',\\
&m_{\zeta}^2 = \V' + \frac{4 \rho_1}{N} \W,  
\end{align}
\end{subequations}
where $\rho_1 = \tr(\Phi^{\dag} \Phi)$, $\V' = \partial \V(\rho_1)/\partial \rho_1$, and
\begin{align}
     \W = \W(\rho_1) \equiv  \left.\frac{\partial V_k(\rho_1, \rho_2, \rho_3, \dots)}{\partial \rho_2}\right|_{\rho_2 = 0, \rho_3=0, \dots}.
\end{align}
Inserting \cref{eq:bosongama2} into \cref{eq:weq}, we obtain
\begin{align}
    \dot{\V}^{(\mathcal{B})} = \frac{1}{2} \int_Q \left[\tr G_k^{(S)} + \tr G_k^{(V)} \right] \partial_t R^{(\mathcal{B})}_k,
\end{align}
with
\begin{equation}
\begin{split}
\label{eq:g2_inverse}
  &G_k^{(S)} =  \left(\bar{\Gamma}^{(S)}_k  + R^{(\mathcal{B})}_k \id_2  \right)^{-1},\\[1ex]
  &G_k^{(V)} = \left(\bar{\Gamma}^{(V)}_k + R^{(\mathcal{B})}_k \id_{2 n_{\zeta}} \right)^{-1}.
\end{split}
\end{equation}
For our choice of the cutoff, \cref{eq:litimb}, one can explicitly perform the momentum integration in $d$ spatial dimensions and the Matsubara summation
\begin{align}
 \label{eq:uflow}
\dot{\V}^{(\mathcal{B})}  &=  C_d  \Z k^{d+2} \left\{ 1 - \frac{ \eta }{d+2} \right\} \\[1ex]   \notag
 &\times \Big(  (\E_{\alpha}+\E_{\sigma})   \mathcal{B}^{(1)}_T(\E_{\alpha} \E_{\sigma}) +   n_{\zeta}(\E_{\pi}+\E_{\zeta})   \mathcal{B}^{(1)}_T(\E_{\pi}\E_{\zeta}) \Big).
\end{align}
Here 
\begin{equation}
\label{eq:cd}
C_d = \frac{1}{(4\pi)^{d/2}\Gamma(d/2+1)}, \quad C_3 = \frac{1}{6\pi^2}, 
\end{equation}
and $\eta = -\partial_t \ln \Z$. The thermal $\mathcal{B}$-functions are defined in \cref{app:matsubara} and depend on 
\begin{equation}
\begin{split}
\label{eq:a_coeff}
    &\E_{\alpha} = \E_{\pi} = \Z k^2+m_{\pi}^2, \\ 
    &\E_{\sigma} = \Z k^2+m_{\sigma}^2, \\
    &\E_{\zeta} = \Z k^2+m_{\zeta}^2.
\end{split}
\end{equation}
The asymptotic behavior of the thermal functions determines two limiting forms of the flow equation. In the high-temperature limit, the bosonic contribution behaves as  $\mathcal{B}^{(1)}_{T\to \infty}(x) \sim T/x$, reflecting that the leading contribution originates solely from the zero Matsubara frequency, $q_0 = 0$. In this regime, the flow equation reduces to that of a classical $\mathrm{U}(N)$-symmetric theory, consistent with the results of \cite{kalagov2023}. In contrast, taking the zero-temperature limit yields  $\mathcal{B}^{(1)}_{T\to 0}(x) \sim 1/(\A \sqrt{x})$, which matches the result one would obtain by formulating the model directly at $T = 0$. However, in this case, the integration over the ``time'' interval $[0, \beta]$ must be extended to $[-\beta/2, \beta/2]$, taking into account the periodic properties of the bosonic field. Only after this extension can one safely take the limit $T\to 0$, since the theory on the time semi-axis differs significantly from the theory on the entire axis.

The equation \cref{eq:uflow} is not closed, as it includes the independent function $\W$ via the mass of the $\zeta$-modes. Note that for $N = 2$, the $\zeta$-subspace has dimension $n_{\zeta} = 0$, making the equation closed and identical to that of the standard $U(1)$ symmetric model.  This follows from the fact that a complex antisymmetric $2\times2$ matrix  is parameterized by a single complex number. For enlarged symmetry ($N \geq 4$), the flow of the derivative of the potential $V_k$ with respect to the $i$-th invariant includes contributions from higher-order invariants. This results in a multidimensional system of partial differential equations. To simplify the analysis, this system must be truncated, retaining only the most relevant part. In this paper, in addition to the evolution of the potential $\V$ itself, we fully account for the flow of $\W$. To incorporate the contributions of the derivatives of the potential $V_k$ with respect to higher-order invariants $\rho_3,\rho_4,\dots$, we take their classical values, which vanish according to \cref{eq:initial_cond}. To calculate the flow for $\W$, one can infinitesimally extend the background $\Phi$ in a certain direction such that $\rho_2 \neq 0$ and  use then the
approximate expression $V_k = \V+ \W \rho_2$. In this way, the coefficient of the linear term in the expansion of the right-hand side of \cref{eq:weq} determines the desired flow for $\W$.  For convenience, one extends the vacuum expectation value $\Phi$ in the orthogonal direction  $\Phi + \Phi_{\bot}$ such that $\tr(\Phi \Phi_{\bot}) = 0$ with
\begin{equation}
\label{eq:delta_bot}
\Phi_{\bot}   = \Delta_{\bot} \begin{pmatrix}
0 & 1 & 0 &  \dots & 0 \\ 
-1 & 0 & 0 & \dots & 0 \\ 
0& 0& 0&  \dots & 0 \\
\dots & \dots & \dots & \dots & 0 \\ 
0 & 0 & 0& \dots & 0
\end{pmatrix},
\end{equation}
where $\Delta_{\bot}$ denotes a small real amplitude. The new background yields two nonvanishing invariants: $\rho_1 = N  \Delta^2 + 2 \Delta_{\bot}^2$ and $\rho_2 = 4 \Delta^2 \Delta_{\bot}^2 + \mathcal{O}(\Delta_{\bot}^4)$, as well as a modified mass spectrum that becomes more complex, see \cref{sec:mass}. Ultimately, we arrive at the flow equation
\begin{align}
\label{eq:wflow} 
\notag
\dot{\W}^{(\mathcal{B})} &= C_d  \Z k^{d+2} \left\{ 1 - \frac{\eta }{d+2} \right\} \\[1ex] \notag 
& \times \Big(  w^{(1)}_{\pi \sigma}   \mathcal{B}^{(1)}_T(\E_{\pi} \E_{\sigma})+ w^{(1)}_{\pi \zeta} \mathcal{B}^{(1)}_T(\E_{\pi} \E_{\zeta}) \\[1ex] \notag
&+ w^{(2)}_{\pi \sigma}   \mathcal{B}^{(2)}_T(\E_{\pi} \E_{\sigma})+ w^{(2)}_{\pi \zeta}  \mathcal{B}^{(2)}_T(\E_{\pi} \E_{\zeta})\\[1ex] 
&+ w^{(3)}_{\pi \zeta}  \mathcal{B}^{(3)}_T(\E_{\pi} \E_{\zeta}) \Big). 
\end{align}
Here, $w$-coefficients depend on the potential and its derivatives, as given in \cref{sec:w_coeff}.

The fermionic loop contribution is computed in a similar manner. It is important to note that, by definition, the fermionic sector contributes to the supertrace $\STr$ with an overall negative sign ``$-1$'', \cref{eq:str}. The Hessian matrix obtained from the second functional derivative with respect to the Grassmann fields is
\begin{align}
\label{eq:g2_fermi}
    [\Gamma_k^{(2)}]^{(\mathcal{F})} = \frac{\overrightarrow{\delta}}{\delta \Psi^{T}} \Gamma_k \frac{\overleftarrow{\delta}}{\delta \Psi}.
\end{align}
The fermionic contribution to the flow of the potential $V_k$ is given by the expression
\begin{align}
\label{eq:fermi_part}
    \dot{V_k}^{(\mathcal{F})} = - \int_Q \tr\left[ \big(q_0^2 +  \xi_{\q, k}^2 \big) \id_N+ g_k^2 \Phi^{\dag} \Phi\right]^{-1} \partial_t R^{(\mathcal{F})}_k,
\end{align}
where $ \xi_{\q, k} = \xi_{\q} + R^{(\mathcal{F})}_k$ and $q_0 \in \pi T (2 \mathbb{Z}+1)$. By considering this equation in the extended background and selecting the zero- and linear-order terms in the expansion with respect to $\rho_2$, we obtain fermionic contributions to \cref{eq:uflow} and \cref{eq:wflow}, respectively
\begin{subequations}
\label{eq:fermi_flow} 
\begin{align}
\label{eq:u_fermi_flow}
    \dot{\V}^{(\mathcal{F})} &= -2 N C_d k^2  \mathcal{F}^{(1)}_{T}(a_{\psi}^2)\\  \notag &
     \times \Big(\theta(\mu+k^2) (\mu+k^2)^{\frac{d}{2}}- \theta(\mu-k^2) (\mu-k^2)^{\frac{d}{2}}\Big), \\[1.5ex]
    \label{eq:w_fermi_flow}
    \dot{\W}^{(\mathcal{F})} &= -2  C_d k^2  g_k^4 \mathcal{F}^{(3)}_{T}(a_{\psi}^2) \\ \notag &
    \times \Big(\theta(\mu+k^2) (\mu+k^2)^{\frac{d}{2}}- \theta(\mu-k^2) (\mu-k^2)^{\frac{d}{2}}\Big).
\end{align}
\end{subequations}
The fermionic thermal $\mathcal{F}$-functions are defined in \cref{app:matsubara} and depend on $a_{\psi}^2 = k^4 + g_k^2 \rho_1/N$. 

\subsection{Flow of the derivative expansion coefficients}

Within the approximation employed here, the difference in the stiffness coefficient $\Z$ between the radial and Goldstone modes is negligible. To evaluate the flow of $\Z$, we compute the renormalization factor associated with the Goldstone $\alpha$-component corresponding to the transverse part of $\Gamma_k^{(2)}$, which excludes radial contributions. For this purpose, we consider a nonuniform ground state constructed by an infinitesimal extension of the homogeneous background along the $\alpha$-direction
\begin{equation}
\label{eq:p_vev}
\phi_{\p}(X) =  \exp\left(\frac{i \alpha_\p(\x)}{ \sqrt{2 N } \,\Delta}\right)\Phi \approx \Phi + \frac{i \alpha_\p(\x)}{\sqrt{2N}}  \idj.
\end{equation} 
Here $\alpha_\p(\x) =   {\mathtt a} \, e^{- i \p \x}  + {\mathtt a}^* e^{i \p \x}$, where ${\mathtt a} \in \mathbb{C}$ is a small amplitude. Substituting this configuration into the ansatz and expressing the factor $\Z$ in terms of the appropriate derivatives of $\Gamma_k$, one obtains
\begin{equation}
\label{eq:z1}
 \partial_t \Z = \vol_{\beta}^{-1} \lim_{\p\to 0} \frac{\partial}{\partial \p^2} \lim_{{\mathtt a} \to 0}  \frac{\partial^2}{\partial{\mathtt a} \partial{\mathtt a}^* } \partial_t \Gamma_k[\phi_{\p}].
\end{equation}
To evaluate the flow of  $\A $, it is necessary to consider explicitly the cross-term  $\phi^{\dag}\partial_0 \phi$, which couples the imaginary and real components of the field. In this case, we adopt the non-uniform ground state in the form 
 \begin{align}
 \label{eq:p0_vev}
 \notag
      \phi_{p_0}(X )   =&  \exp\left(\frac{i \alpha_{p_0}(x_0)}{ \sqrt{2 N } \,\Delta}\right)   \Big( \Phi +\frac{\sigma_{p_0}(x_0) }{\sqrt{2 N }} \idj\Big)\\[1ex] \approx & \Phi  +\frac{\sigma_{p_0}(x_0) + i \alpha_{p_0}(x_0)}{\sqrt{2N}} \idj,
 \end{align}
where now $\alpha_{p_0}(x_0) =   {\mathtt a} e^{i p_0 x_0}  + {\mathtt a}^* e^{-i  p_0 x_0} $ and $\sigma_{p_0}(x_0) =  {\mathtt s} e^{ i  p_0 x_0}  + {\mathtt s}^* e^{-i  p_0 x_0}$. This leads to an expression structurally analogous to \cref{eq:z1}
\begin{equation}
\label{eq:z2}
\partial_t  \A  =  \vol_{\beta}^{-1} \lim_{p_0 \to 0} \frac{\partial}{\partial p_0}  \lim_{{\mathtt a}, {\mathtt s}\to 0}  \frac{\partial^2}{\partial{\mathtt s}\partial {\mathtt a}^* } \partial_t \Gamma_k[\phi_{p_0}]. 
\end{equation}
A direct computation (see \cref{app:z_flow}) yields the bosonic part of the flow equations
\begin{widetext}
\begin{subequations}
\label{eq:der_coeff_flow_b} 
\begin{align}
 \dot{\Z}^{(\mathcal{B})}  &=  -4 \rho_{1}  C_d \Z^2 k^{d+2}\Big( \lambda^2_{\sigma}   \mathcal{B}^{(2)}_T(\E_{\pi} \E_{\sigma} ) + n_{\zeta} \lambda^2_{\xi}  \mathcal{B}^{(2)}_T( \E_{\pi} \E_{\zeta})\Big),\\[1.5ex] 
 \dot{\A}^{(\mathcal{B})}   &=  -2 \rho_{1}  C_d \Z \A k^{d+2} \left\{ 1 - \frac{\eta }{d+2} \right\}  \Big( \lambda^2_{\sigma}  (b_{\sigma}-1)   \mathcal{B}^{(2)}_T(\E_{\pi} \E_{\sigma}) + n_{\zeta} \lambda^2_{\zeta}  (b_{\zeta}-b_{\pi})   \mathcal{B}^{(2)}_T(\E_{\pi}\E_{\zeta} )  \\[1ex] \nonumber
  & - 2 \lambda^2_{\sigma} (b_{\sigma} \E_{\pi}-\E_{\sigma} ) ( \E_{\pi}+\E_{\sigma})   \mathcal{B}^{(3)}_T(\E_{\pi} \E_{\sigma})     - 2  n_{\zeta} \lambda^2_{\zeta}  (b_{\zeta} \E_{\pi}-b_{\pi} \E_{\zeta} ) ( \E_{\pi}+\E_{\zeta})  \mathcal{B}^{(3)}_T(\E_{\pi} \E_{\zeta})\Big).
\end{align}
\end{subequations}
\end{widetext}
The projection of the fermionic part of the supertrace $\STr$ onto the background fields specified in \cref{eq:p_vev,eq:p0_vev}  yields 
\begin{subequations}
\label{eq:der_coeff_flow_f} 
  \begin{align}
  &\dot{\Z}^{(\mathcal{F})}   = - C_d k^{d-4}  g_k^2 \mathcal{F}^{(2)}_{\bar{T}}(\bar{\E}_{\psi}^2) \\ \notag
    &\times \Big( \theta(\bar{\mu}+1) (\bar{\mu}+1)^{\frac{d}{2}}+ \theta(\bar{\mu}-1) (\bar{\mu}-1)^{\frac{d}{2}}\Big), \\[1.5ex] 
    &\dot{\A}^{(\mathcal{F})}   =  C_d k^{d-4}  g_k^2 \Big(\mathcal{F}^{(2)}_{\bar{T}}(\bar{\E}_{\psi}^2) - 4 \mathcal{F}^{(3)}_{\bar{T}}(\bar{\E}_{\psi}^2)\Big) \\ \notag
    &\times \Big( \theta(\bar{\mu}+1) (\bar{\mu}+1)^{\frac{d}{2}}+ \theta(\bar{\mu}-1) (\bar{\mu}-1)^{\frac{d}{2}}  - 2 \bar{\mu}^{\frac{d}{2}}\Big),    
 \end{align}
 \end{subequations}
 where $\bar{\mu} = \mu/k^2$, $\bar{T} = T/k^2$, $\bar{\E}_{\psi}^2 = 1+ g_k^2 \rho_{1}/ N k^4$. 
Note that all quantities in \cref{eq:der_coeff_flow_b,eq:der_coeff_flow_f} are evaluated at the running minimum of the potential $\rho_1 = \rho_{0, k}$  defined by $\V'(\rho_{0,k}) = 0$, and  $\lambda_{\sigma} \equiv \V''(\rho_{0,k})$, $\lambda_{\zeta} \equiv 2\W(\rho_{0,k})/N$.

\section{\label{sec:results}Results}

To obtain the global solution of the flow equations, we apply the method of lines. Specifically, the functions $\V(\rho_1)$ and $\W(\rho_1)$ are discretized on a grid in terms of the gap variable $\Delta = \sqrt{\rho_1/N}$ using a finite-difference scheme, as detailed in Appendix~\ref{app:numeric}. The resulting high-dimensional system of ordinary differential equations is then solved using an adaptive Runge--Kutta--Fehlberg method, which ensures numerical stability and accuracy. Direct numerical evaluation of observables at $k = 0$ is not feasible. However, the renormalization group flow typically freezes out at a finite scale $k_0$, well below the characteristic physical scales of the system such as the thermal de~Broglie wavelength $\sim \sqrt{T}$ and the scale $\sim \sqrt{\mu}$. To  extract the phase structure,  we terminate the integration of the flow equation at $k_0$. In our computations, we choose $k_0 \approx 0.02\sqrt{\mu}$, ensuring that the flow has effectively frozen out—that is, observables no longer exhibit significant scale dependence within the limits of numerical accuracy. To ensure consistency, the initial UV scale $\Lambda$ must be taken sufficiently large to suppress many-body effects at the onset of the renormalization group flow. We set this scale to $\Lambda = 200 \sqrt{\mu}$ and verified that the numerical results remain stable upon further increases of $\Lambda$.

\begin{figure}
\includegraphics[width=\linewidth]{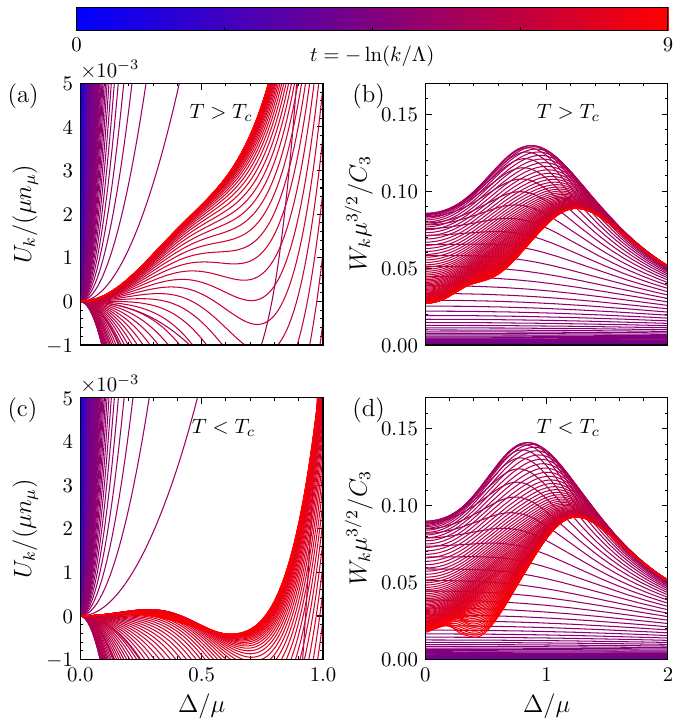}
\caption{(Color online). Scale dependence of the running potential $\V$ and the coefficient of the invariant expansion $\W$. (a) Above the critical temperature, the RG flow drives the potential toward a limiting shape (shown as the red limit curve), characterized by a trivial global minimum at the origin. (b,d) The function $\W$, initialized as a constant vanishing at the UV scale (blue/violet region), develops a nontrivial structure both above and below the critical temperature.
(c) At low temperatures, the RG flow evolves the potential into a double-well structure, where a nontrivial minimum emerges discontinuously, signaling a first-order phase transition. In this figure and those that follow,  we define $n_{\mu}  \equiv N C_3 \mu^{3/2}$. }
\label{fig:flow}
\end{figure}

\begin{figure}
\includegraphics[width=\linewidth]{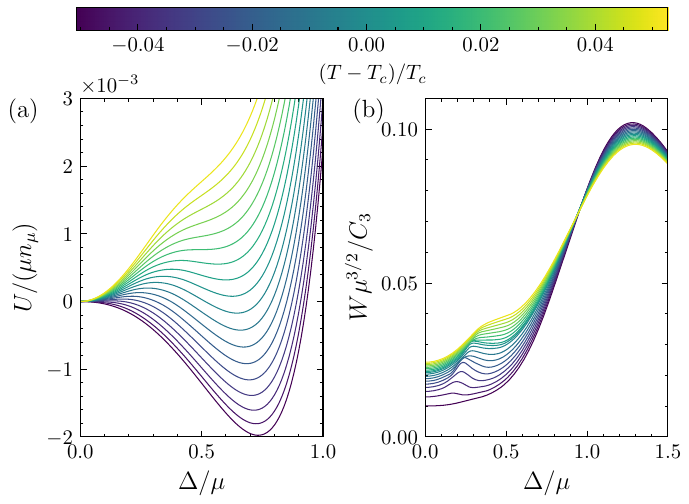}
\caption{(Color online). (a) Temperature dependence of the full effective potential $U \equiv U_{k_0}$. (b) For completeness, the shape of the function  $W \equiv W_{k_0}$ is also shown.}
\label{fig:ir_flow}
\end{figure} 

A typical evolution of the running potential from the UV scale $k = \Lambda$ to the IR limit $k = k_0$, for $N\geq 4$, is shown in \cref{fig:flow}. Bosonic fluctuations, consistently incorporated via the renormalization group equations, induce a first-order phase transition. In contrast, retaining only fermionic degrees of freedom reproduces the well-known mean-field results discussed in \cref{sec:mf}, regardless of the number of spin components $N$.  This is confirmed explicitly  by integrating the fermionic part of the flow equation \cref{eq:fermi_part}.  The location of the minimum $\Delta_c$ of the potential  and the temperature $T_c$ at which the emerging nontrivial minimum becomes the global one, indicating the transition to the ordered phase, are determined via standard thermodynamic relations. These include the extremum condition $ \partial_{\Delta} U_{k_0}(\Delta)|_{\substack{\Delta = \Delta_c, T = T_c}} = 0$, and the requirement that both minima have equal free energy $U_{k_0}(0) =  U_{k_0}(\Delta_c)$ at $T = T_c$.

\begin{figure}
\includegraphics[width=\linewidth]{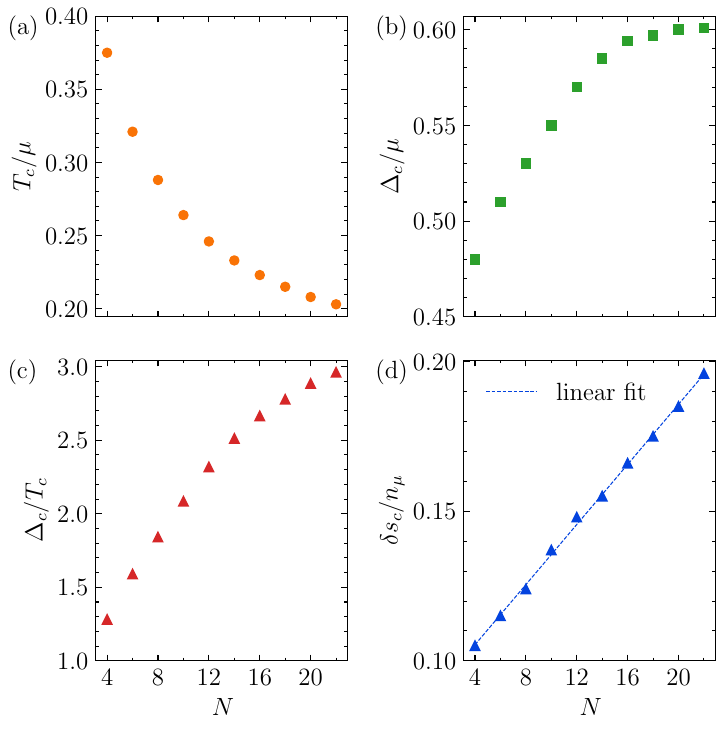}
\caption{(Color online). Thermodynamic parameters of the first-order phase transition as functions of  the number of spin components $N$. (a) Critical temperature $T_c/\mu$; (b) energy gap at the critical temperature $\Delta_c/\mu$; (c) ratio of the gap magnitude to the critical temperature  $\Delta_c/T_c$; (d) discontinuity in the entropy density at the transition point $\delta s_c/n_{\mu}$. The numerical data in panel (d) (blue triangles) are well approximated by a linear fit (blue dashed line) of the form $f(x) = 0.005 x+ 0.085$. The corresponding numerical values are listed in \cref{tab:thermo_quantities}.}
\label{fig:n_depend}
\end{figure}

Shown in \cref{fig:ir_flow} are the computed results  for the final form $U_{k_0}(\Delta)$ of the effective average potential at different  temperatures. The full system of renormalization group equations was solved using a uniform temperature step of $10^{-3}\mu$, for fixed values of $N$. The quantitative results are summarized in \cref{tab:thermo_quantities} and visually represented in \cref{fig:n_depend}. The temperature of the first-order phase transition decreases monotonically with increasing $N$, starting from its maximum value $T_c/\mu\approx 0.375$ at $N=4$---the smallest number of spin components for which the system exhibits a discontinuous transition. This behavior is consistent with thermodynamic expectations: as $N$ increases, the entropy of the normal phase rises due to the growing number of accessible internal configurations, which favors the disordered state and suppresses the formation of superfluid order. Although the critical temperature continues to decrease up to the highest spin multiplicity examined ($N = 22$), we anticipate it to eventually reach a plateau, as beyond a certain value of  $N$, the entropic advantage of the normal phase no longer significantly suppresses the formation of the ordered phase. By contrast, the discontinuity in the order parameter, quantified by the energy gap at the transition point  $\Delta_c/\mu$, increases monotonically with $N$ and  approaches a finite limiting value for $N \gg1 $; see \cref{fig:n_depend}(b). The increase in entropy necessitates a larger pairing amplitude to stabilize the superfluid phase via a first-order transition, but as $N$ grows further, this effect diminishes, leading to the saturation of the gap discontinuity. In this sense, one may say that fluctuations are enhanced by a larger number of spin components, thereby making the first-order phase transition more pronounced. The discontinuous nature of the identified phase transition allows us to introduce a new characteristic quantity: the ratio of the energy gap at the transition point to the critical temperature $\Delta_c/T_c$,  \cref{fig:n_depend}(c). This invites an analogy with the continuous  transition in the classical $\mathrm{SU}(2)$ BCS model, where the ratio of the zero-temperature energy gap to the critical temperature is known to be universal. 

\begin{table}
\centering
\caption{Thermodynamic characteristics of the first-order phase transition  as functions of the spin multiplicity $N$. The table lists the critical temperature $T_c/\mu$, the energy gap at the transition point $\Delta_c/\mu$ (with $\mu$ evaluated at the critical temperature), and the ratio $\Delta_c / T_c$. The estimated numerical uncertainties are  $1 \%$ for $T_c$ and $3$--$4\%$ for $\Delta_c$; see \cref{app:numeric} for details.}
\setlength{\tabcolsep}{10pt} 
\begin{tabular}{lSSS}
\hline\hline
\( N \) & \( T_c/\mu \) & \( \Delta_c/\mu \) & \( \Delta_c / T_c \) \\
\hline
4  & 0.375 & 0.48 & 1.28 \\
6  & 0.321 & 0.51 & 1.59 \\
8  & 0.288 & 0.53 & 1.84 \\
10 & 0.264 & 0.55 & 2.10\\
12 & 0.246 & 0.57 & 2.32 \\
14 & 0.233 & 0.585 & 2.51 \\
16 & 0.223 & 0.59 & 2.66 \\
18 & 0.215 & 0.597 & 2.78 \\
20 & 0.208 & 0.60 & 2.88 \\
22 & 0.203 & 0.60 & 2.95 \\
\hline\hline
\end{tabular}
\label{tab:thermo_quantities}
\end{table}

\begin{figure}[t]
\includegraphics[scale=1]{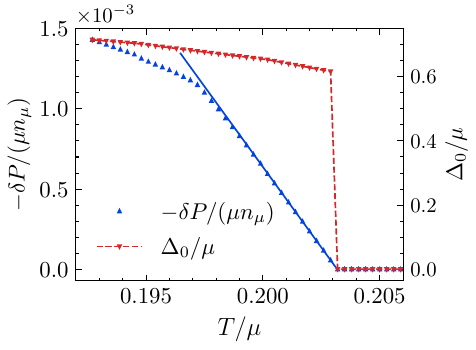}
\caption{(Color online). Typical  temperature dependence of the pressure change (blue upward triangles) and the superfluid gap (red  downward triangles) in the vicinity of the first-order phase transition.  The solid blue line highlights the pronounced slope of the pressure curve. Below $T_c$ (at  $T/\mu \approx 0.197$ in the case depicted), the pressure also appears to change noticeably---at least visually, within the given image resolution. However, the derivatives of the pressure remain continuous in this region, and no anomalies are observed.}
\label{fig:press}
\end{figure}

The minimum value of the effective potential at fixed $T$ and $\mu$ defines the pressure in the system via $P= - \Omega/ \vol = - U_{k_0}(\Delta_0)$, where the discontinuous behavior of $\Delta_0$ as a function of temperature near the phase transition point is illustrated in \cref{fig:press}. The grand potential is determined up to an additive term, which is fixed by requiring proper vacuum behavior. However, within the present approach, it is not possible to reliably compute the full zero-temperature potential $V_k$, since it is a non-analytic function, rendering the expansion in field invariants inapplicable.  Instead, we analyze the change of the thermodynamic variable across the finite temperature phase transition. We define the pressure change as $\delta P \equiv  P_{+} - P_{-}$, where $ P_{+}$ is the pressure in the disordered phase $T>T_c$ and is given by the trivial minimum $P_{+} = -U_{k_0}(0)$, while $P_{-}$ is computed below $T_c$, thus $P_{-}|_{T = T_c}= -U_{k_0}(\Delta_c)$. While the pressure remains continuous at $T=T_c$, $\delta P(T_c ) = 0$, it exhibits a pronounced slope just below $T_c$, as seen in \cref{fig:press}, which is associated with the entropy discontinuity. Indeed,  the entropy per unit volume is given by  $s = (\partial P/\partial T)_{\mu}$, thus its change defined as $\delta s = (\partial P_{+}/\partial T)_{\mu} - (\partial P_{-}/\partial T)_{\mu}$ is a discontinuous function, such that the right-side limit is $\delta s(T_c+0) = 0$, whereas the left-sided limit is nonzero: $\delta s_c \equiv \delta s(T_c-0) \neq 0$. When the energy gap opens discontinuously at the critical temperature, low-energy excitations are abruptly suppressed. As a result, the number of thermally accessible microstates decreases, leading to a sudden drop in entropy. The behavior of the entropy jump $\delta s_c$ as a function of spin multiplicity  is shown in \cref{fig:n_depend}(d). We find that this thermodynamic quantity exhibits a simple linear dependence on $N$. As $N$ increases, the system gains additional internal degrees of freedom associated with its $\mathrm{SU}(N)$ spin symmetry. Upon undergoing the phase transition to the superfluid state, a pairing gap opens for all components, resulting in a more pronounced suppression of low-energy excitations.  Consequently, the reduction in the number of accessible microstates at the transition becomes more significant with larger $N$, and the entropy drop grows accordingly.

Despite the absence of a direct analogue of $\delta s_c/n_{\mu}$  in two-component systems, we may compare it to the entropy per particle $S_c/ \mathscr{N}$ at the continuous phase transition point in unitary $\mathrm{SU}(2)$ gases. For a rough estimate, we assume that the chemical potential is of the order of the Fermi energy,  $\mu \sim  \varepsilon_F = k_F^2$, where the Fermi momentum is defined as $k_F = (\mathscr{N} / N \vol C_3)^{1/3}$.  Under this assumption, the entropy jump per particle is approximately $\delta S_c/\mathscr{N} \sim \delta s_c/n_{\mu} \approx 0.1 - 0.2$, as shown in \cref{fig:n_depend}(d). 
The predicted reference value of $S_c/\mathscr{N}$ that most closely matches the experimental data \cite{Ku2012} is  $S_c/\mathscr{N} \approx 0.71$, as obtained in \cite{Haussmann2007}. A comparison of the thermodynamic signatures of the discontinuous transition in the $\mathrm{SU}(N \geq4)$ system, \cref{tab:thermo_quantities}, and some similar observables in the two-component case, \cref{tab:thermo_quantities_su2}, reveals a numerical proximity---at least at the level of order of magnitude. This suggests that the first-order nature of the established phase transition does not make it weak or experimentally elusive. Therefore, the transition should be accessible in realistic experimental conditions, providing a clear and measurable manifestation of many-body effects and symmetry-enhanced pairing in fermionic systems beyond the conventional $\mathrm{SU}(2)$ paradigm.

\begin{table}
\caption{Critical temperature $T_c/\mu$ and zero temperature energy gap $\Delta/\mu$ for the $\mathrm{SU}(2)$ system (in the unitary limit), which undergoes a continuous phase transition.}
\label{tab:thermo_quantities_su2}
\centering
\begin{tabular}{lSS}
\hline\hline
Method & {$T_c/\mu$} & {$\Delta/\mu$}  \\
\hline
Experiment \cite{Ku2012}                  & 0.40         &      \\
FRG \cite{Boettcher14}                     & 0.38(2)      & 1.04(15) \\
Luttinger--Ward formalism \cite{Haussmann2007} & 0.40         & 1.27    \\
Quantum Monte Carlo \cite{Burovski2006}     & 0.31(2)      &     \\
Mean field [\cref{sec:mf}]                             & 0.66         & 1.16    \\
\hline\hline
\end{tabular}
\end{table}

\section{\label{sec:consl}Conclusions and outlook}

Our analysis is based on the functional renormalization group approach, which enables a nonperturbative computation of the thermodynamic potential across a broad range of physical parameters, properly accounting for both quantum and thermal fluctuations. We applied this technique to  study  superfluidity in a system of attractive large-spin Fermi particles with an $\mathrm{SU}(N)$-symmetric pairwise interaction. To this end, we computed the effective potential at leading order in the derivative expansion of the partially bosonized effective average action. To obtain the full functional form of the potential, we discretized the flow equations on a nonuniform grid and solved them using  an adaptive Runge--Kutta scheme.

Our results highlight the essential role of bosonic degrees of freedom in the large-spin case ($N \geq 4$). While the mean-field approximation---based solely on fermionic contributions---predicts a continuous phase transition, the inclusion of the order-parameter field fluctuations leads to the first-order transition. Notably, in the well-known $\mathrm{SU}(2)$ case, the FRG analysis yields only corrections to the critical exponents and other thermodynamic observables without qualitatively altering the nature of the transition compared to the naive mean-field theory.

We presented quantitative results for the key characteristics of the established first-order phase transition, including the critical temperature and the discontinuities in both the superfluid gap and the entropy density, in the unitary regime ($1/a_s \to 0^{-}$) for $N \geq 4$. We found that the critical temperature $T_c/\mu$ decreases with increasing spin multiplicity $N$,  whereas the energy gap $\Delta_c/\mu$ and the entropy jump $\delta s_c/n_{\mu}$ increase. While former saturates at large $N$, the latter shows a linear dependence on $N$. 

In this work, we did not address the determination of the equation of state or the expression of thermodynamic quantities in terms of the Fermi energy. We also did not explore potential refinements of the computed results that could arise from extending the truncation of the effective action or employing a frequency-dependent regulator. The technical framework presented here can  be extended to address these issues, although doing so would require more intensive numerical computations. This provides  natural directions for future investigations of systems with tunable $\mathrm{SU}(N)$ symmetry.

\begin{acknowledgments}
We thank M. Hnati\v{c}, M. Nalimov and A. Sutulin for inspiring discussions. The work was supported by the Foundation for the Advancement of Theoretical Physics and Mathematics ``BASIS''
\end{acknowledgments}

\appendix

\section{\label{app:matsubara}Matsubara sums}

This Appendix presents the thermal functions that arise in the flow equations, defined through the Matsubara sums. In the fermionic sector, the sums are taken over frequencies $q_0 \in  \pi T (2 \mathbb{Z}+1)$ and are given by the expression

\begin{align}
    \mathcal{F}^{(l)}_T(x) & = T \sum_{q_0 } \big(x+ q_0 ^2\big)^{-l}.
\end{align} 
They are calculated recursively 
\begin{align}
     \mathcal{F}^{(l)}_T(x) = \frac{(-1)^{l-1}}{(l-1)!}\frac{d^{l-1}}{dx^{l-1}}   \mathcal{F}^{(1)}_T(x),
\end{align}
where 
\begin{align}
\notag
        \mathcal{F}^{(1)}_T(x) &=  \frac{1}{2 \sqrt{x}} \big(1-2  N^{(\mathcal{F})}_T(\sqrt{x})\big),\\ 
      N^{(\mathcal{F})}_T(x) &= \frac{1}{\exp(x/T)+ 1}. 
\end{align}
In the bosonic parts of the flow equations, the sums are taken over frequencies $q_0 \in  2 \pi T \mathbb{Z}$ and are defined by the expression
\begin{align}
   \mathcal{B}^{(l)}_T(x) & = T \sum_{q_0 } \big(x+ \A^2 q_0 ^2\big)^{-l}, 
\end{align} 
with
\begin{align}
     \mathcal{B}^{(l)}_T(x) = \frac{(-1)^{l-1}}{(l-1)!}\frac{d^{l-1}}{dx^{l-1}}  \mathcal{B}^{(1)}_T(x),
\end{align}
where 
\begin{align}
\notag
    \mathcal{B}^{(1)}_T(x) &=  \frac{1}{2 \A \sqrt{x}} (1+2 N^{ (\mathcal{B})}_T(\sqrt{x}/\A)),\\ 
     N^{ (\mathcal{B})}_T(x) &= \frac{1}{\exp(x/T) - 1}.
\end{align}

\section{\label{sec:mass} Diagonalization}
In this section, we show how to evaluate the bosonic part of the supertrace $\STr$ in \cref{eq:weq} on the extended background. For simplicity, we introduce the notation  $M \equiv [\Gamma_k^{(2)}]^{(\mathcal{B})}|_{\phi = \Phi + \Phi_{\bot}}$.  It can be expanded in powers of $\rho_2$  as follows 
\begin{equation}
\label{eq:M_expansion}
 M =  M^{(0)} + M^{(1)}  \sqrt{\rho_2} + M^{(2)} \rho_2+\mathcal{O}(\rho^{3/2}). 
\end{equation}
To the same order of accuracy, we can write
\begin{align}  
    \Tr \ln(M+R_k) &\approx  \ \Tr  \ln(G_0^{-1}) +  \Tr \left(G_0 M^{(1)} \right)\sqrt{\rho_2} \\ \notag
    &  +  \Tr  \left(G_0M^{(2)}  - \frac{1}{2} G_0 M^{(1)} G_0M^{(1)}\right) \rho_2,
\end{align}
where $G_0^{-1} =M^{(0)}+R_k =(\Gamma_k^{(2)}[\phi]+R_k)|_{\phi = \Phi}$.  The leading term in this expansion yields the flow equation for the potential $U$
\begin{align}
   \dot{\V}^{(\mathcal{B})}  = \frac{1}{2} \hat{\partial}_t \Tr  \ln( M^{(0)}+R_k),
\end{align}
which can be straightforwardly computed in the basis defined in \cref{eq:nonlin}. Looking ahead, we note that the second term $\sim \sqrt{\rho_2}$ vanishes due to the inherent symmetry of the model—a fact that can be explicitly confirmed through direct calculation. The third term provides the desired flow equation for the function $W$

\begin{align}
\label{eq:w}
    \dot{\W}^{(\mathcal{B})} = \frac{1}{2}\hat{\partial}_t \Tr  \left(G_0M^{(2)}  - \frac{1}{2} G_0 M^{(1)} G_0M^{(1)}\right).
\end{align}
The representation given in \cref{eq:nonlin} is convenient for evaluating $M^{(0)}$ in the case of symmetry breaking $\mathrm{U}(N) \to \mathrm{USp}(N)$. However, the presence of a nonzero additive term $\Delta_{\bot}$, \cref{eq:delta_bot}, alters this pattern, prompting us to adopt an alternative parametrization of the field $\phi$. In the new basis, the full matrix $M$ takes a block-diagonal form, with each square diagonal block being irreducible; consequently, the trace in \cref{eq:w} decomposes into the sum of those over the corresponding subspaces. To this end, we split the fluctuating part of the  field into its real and imaginary components, $\phi = \phi^{\re} + i \phi^{\im}$. The real antisymmetric matrices $\phi^{\re}$ and $\phi^{\im}$ are represented as follows

\begin{equation}
\label{eq:decompos}
\phi^{\re} = \begin{pmatrix}
 A^{\re} + C^{\re} & B^{\re}_a + B^{\re}_s \\ 
 B^{\re}_a-B^{\re}_s & A^{\re} - C^{\re} \\
 \end{pmatrix} +  \zeta^{\re}_1 \idj, 
\end{equation}
where $A^{\re}, B_a^{\re}, C^{\re}$ are  $(N/2)\times(N/2)$ antisymmetric matrices, while $B^{\re}_s$ denotes a $(N/2)\times(N/2)$ symmetric matrix. The same applies to the matrix $\phi^{\im}$. This representation merely corresponds to a reorganization of the general expansion, \cref{eq:nonlin}.  We now proceed further by expressing each constituent matrix in the following form 

\begin{align}
\notag
A^{\re} &= \begin{pmatrix}
 \zeta^{\re}_2\, \epsilon & A^{\re}_1 \\ 
 -{A^{\re}_1}^{T} & A^{\re}_2 \\
 \end{pmatrix}, \,
C^{\re} = \begin{pmatrix}
 \zeta^{\re}_3 \, \epsilon & C^{\re}_1 \\ 
 -{C^{\re}_1}^{T} & C^{\re}_2 \\
 \end{pmatrix}, \, \\[2ex]\notag
B^{\re}_s &= \begin{pmatrix}
 B^{\re}_0 + \frac{1}{2}\, \zeta^{\re}_4 \, \id_2 & B^{\re}_1 \\ 
 {B^{\re}_1}^{T} & B^{\re}_2 - \frac{2}{N-4}\, \zeta^{\re}_4 \, \id_{N/2-2} \\
 \end{pmatrix},\, \\[2ex]
B^{\re}_a &= \begin{pmatrix}
  \zeta^{\re}_5 \, J & B^{\re}_3 \\ 
 -{B^{\re}_3}^{T} & B^{\re}_4\\
 \end{pmatrix},
\end{align}
where $\epsilon$ is the unit $2 \times 2$ antisymmetric tensor, see \cref{eq:delta_bot}, $A^{\re}_1, C^{\re}_1, B^{\re}_1, B^{\re}_3$ are $2\times (N/2-2)$ matrices. These matrices can be arranged into two-component row representations: $A^{\re}_1 = ( A^{\re}_{11}, A^{\re}_{12} )^T$, made of $(N/2-2)$-dimensional vectors $A^{\re}_{11}$, etc. All of them can be assembled into a single $8(N/2-2)$-dimensional vector 

\begin{align}
    \mathbf{V}^{\re} =  \begin{pmatrix}
  A^{\re}_{11} \\ 
 A^{\re}_{12} \\
 C^{\re}_{11} \\ 
 C^{\re}_{12} \\
  B^{\re}_{11} \\
  B^{\re}_{12} \\
  B^{\re}_{31} \\
  B^{\re}_{32} \\
 \end{pmatrix}.
\end{align}
Similarly, we combine $\zeta^{\re}$-modes in the $5$-dimensional vector 
$\mathbf{S}^{\re} = (\zeta^{\re}_1,\dots,\zeta^{\re}_5)^T$.  The rest fields  $A^{\re}_2$, $C^{\re}_2$ and $B^{\re}_4$ are $(N/2-2) \times (N/2-2)$ antisymmetric matrices, $B^{\re}_0$ and $B^{\re}_2$ are traceless symmetric matrices sizes of which are $2 \times 2$ and $(N/2-2) \times (N/2-2)$, respectively. They form the column 

\begin{align}
    \mathbf{M}^{\re} =  \begin{pmatrix}
  A^{\re}_2 \\ 
 C^{\re}_2 \\
 B^{\re}_4 \\ 
 B^{\re}_0 \\
  B^{\re}_2 \\
 \end{pmatrix}.
\end{align}
Analogous constructions are carried out for the imaginary part $\phi^{\im}$ of the field, leading to $\mathbf{S}^{\im}$, $\mathbf{V}^{\im}$ and  $\mathbf{M}^{\im}$. Then the quadratic in the bosonic fields term of the effective action $\Gamma_k[\phi]$, \cref{eq:lpa},  is given by
\begin{align}
\notag
   [\Gamma_k[\phi]]^{(\mathcal{B})}  &= \frac{1}{2}  \begin{pmatrix}
  \mathbf{S}^{\re} \\ 
\mathbf{S}^{\im} \\
 \end{pmatrix}^{T} 
 M_{\mathbf{S}} \begin{pmatrix}
  \mathbf{S}^{\re} \\ 
\mathbf{S}^{\im} \\
 \end{pmatrix} \\[2ex] \notag
 &+\frac{1}{2}  \begin{pmatrix}
  \mathbf{V}^{\re} \\ 
\mathbf{V}^{\im} \\
 \end{pmatrix}^{T} 
 M_{\mathbf{V}} \begin{pmatrix}
  \mathbf{V}^{\re} \\ 
\mathbf{V}^{\im} \\
 \end{pmatrix}\\[2ex]
&+ \tr \frac{1}{2}  \begin{pmatrix}
  \mathbf{M}^{\re} \\ 
\mathbf{M}^{\im} \\
 \end{pmatrix}^{T} 
 M_{\mathbf{M}} \begin{pmatrix}
  \mathbf{M}^{\re} \\ 
\mathbf{M}^{\im} \\
 \end{pmatrix}.
\end{align}
The integration $\int_X$ is understood to be implicit. The representation obtained above allows the matrix $M$ to be expressed in the block form
\begin{equation}
M = \begin{pmatrix}
 M_{\mathbf{S}}  &   &   \\
  &  M_{\mathbf{V}}  &   \\
  &   &  M_{\mathbf{M}} 
\end{pmatrix}.
\end{equation}
Furthermore, each diagonal block itself has a nested block structure. Two of them are irreducible 
\begin{equation}
\label{eq:ms}
    M_{\mathbf{S}} = \begin{pmatrix}
 m_{\mathbf{s}}^{\re \re}  &  q_0 \A \id_5    \\
  -q_0 \A \id_5 &  m_{\mathbf{s}}^{\im \im}    
\end{pmatrix},
\end{equation}
and
\begin{align}
    \label{eq:mv}
    M_{\mathbf{V}} = \begin{pmatrix}
 m_{\mathbf{v}}^{\re \re} \otimes \id_{N/2-2}   &  q_0 \A\id_{8(N/2-2)}    \\
  -q_0 \A  \id_{8(N/2-2)} &  m_{\mathbf{v}}^{\im \im}   \otimes \id_{N/2-2}   
\end{pmatrix},
\end{align}
while the third is itself block-diagonal 
\begin{equation}
M_{\mathbf{M}} = \begin{pmatrix}
 M_{A_2}  &   &  & &  \\
  &   M_{C_2} &  & &  \\
  &   &   M_{B_4} & & \\
  &   &         &  M_{B_0}  & \\
  &   &         &   & M_{B_2}
\end{pmatrix}.
\end{equation}
It is evident that all elements here exhibit a structure analogous to \cref{eq:ms,eq:mv}, and therefore, we provide only the first component explicitly
\begin{equation}
    M_{A_2} = \begin{pmatrix}
 m_{A_2}^{\re \re}  &  q_0 \A     \\
  -q_0 \A&  m_{A_2}^{\im \im}    
\end{pmatrix} \otimes \id_{(N/2-2)}.
\end{equation}
All expressions for the matrix elements of $M$, accurate up to linear order in $\rho_2$  are provided in the Supplemental Materials \cite{SM}. By extracting the corresponding expansion coefficients from \cref{eq:M_expansion} and subsequently evaluating the traces in \cref{eq:w}, we obtain the flow equation \cref{eq:wflow}, as presented in the main text.

\begin{widetext}

\section{\label{sec:w_coeff} $w$-coefficients}

In this Appendix we present the coefficients of \cref{eq:wflow}:
\begin{subequations}
\label{eq:wcoef} 
\begin{align}
w^{(1)}_{\pi \sigma} &=   (3 b_{\zeta} - 5 b_{\pi} -4  ) \frac{\W}{\rho_1} + 2\rho_1 \W''  +  \left(  \frac{(\E_{\sigma} - \E_{\pi})(\E_{\pi} - \E_{\zeta})}{\E_{\pi}^2} +  (1 + b_{\pi})^2 + (1 - b_{\zeta})^2 \right) \frac{4 \W^2}{N (\E_{\sigma} - \E_{\zeta})},\\[1.5ex]
w^{(1)}_{\pi \xi} &=  -w^{(1)}_{\pi \sigma} + 2 (n_{\zeta}+5) \W' + 2\rho_1 \W'', \\[1.5ex]
  \frac{w^{(2)}_{\pi \sigma}}{(\E_{\pi} + \E_{\sigma}) } &=  \Big((8 + 9 b_{\pi} - 5 b_{\zeta}) \E_{\pi}  +  (b_{\pi} - b_{\zeta}) \E_{\sigma}\Big) \frac{\W
   }{2 \rho_1}-   2 \E_{\pi}  \rho_1 \W'' + (\E_{\sigma} + b_{\pi}^2 \E_{\pi} ) \frac{4\W^2}{N \E_{\pi} } \\[1ex]  \nonumber & - 
    \Big((\E_{\sigma} - b_{\zeta}\E_{\pi})^ 2 + (1 + b_{\pi})^2  \E_{\pi} \E_{\sigma}\Big) \frac{4 \W^2} {N \E_{\pi} (\E_{\sigma}-\E_{\zeta})},\\[1.5ex]
 \frac{w^{(2)}_{\pi \zeta}}{(\E_{\pi} + \E_{\zeta}) } &=  \Big([2(2 n_{\zeta}+5) + (n_{\zeta}+2) b_{\pi} -  (n_{\zeta}+4) b_{\zeta}] \E_{\pi}  +  [2 +  (n_{\zeta}-2) b_{\pi} -  n_{\zeta} b_{\zeta}] \E_{\zeta}  \Big) \frac{\W
   }{2 \rho_1} \\[1ex] \nonumber &- \Big((7 N^2-14 N-58)\E_{\pi}^2 + (3 N^2-6 N-28) \E_{\pi} \E_{\zeta} - 2\E_{\zeta}^2  \Big) \frac{2\W^2}{N \E_{\pi} (\E_{\pi} + \E_{\zeta}) } \\[1ex] \nonumber  & + 
    \Big((\E_{\zeta} - b_{\zeta}\E_{\pi})^ 2 +(1 + b_{\pi})^2  \E_{\pi} \E_{\zeta} \Big) \frac{4 \W^2} {N \E_{\pi} (\E_{\sigma}-\E_{\zeta})},   \\[1.5ex]
 w^{(3)}_{\pi \zeta} &=  \frac{(N-4) (N+2)}{N} (\E_{\pi}+\E_{\zeta}) (3 \E_{\pi}+\E_{\zeta})^2 \W^2. 
\end{align}
\end{subequations}
Here $a$-and $b$-coefficients are defined in \cref{eq:a_coeff} and \cref{eq:b_coeff}, respectively.

\section{\label{app:z_flow}  Calculation of the flows  for $\Z$ and $\A$ }

To calculate the right-hand side of \cref{eq:z1,eq:z2}, we use the decomposition of the Hessian,  $\Gamma_k^{(2)}[\phi_{\p}] =  \Gamma_k^{(2)} +   \widetilde{\Gamma}_{k, \p}^{(2)}$ in \cref{eq:z1} and $\Gamma_k^{(2)}[\phi_{p_0}] =  \Gamma_k^{(2)} +   \widetilde{\Gamma}_{k, p_0}^{(2)}$ in \cref{eq:z2}. In each case, all dependencies on ${\mathtt a}$ and  $({\mathtt a},{\mathtt s})$ are grouped in the second term. The unperturbed Hessian  $ \Gamma_k^{(2)}$, \cref{eq:unperturb_hess},  is given by the uniform part of the background  configuration. Using these decompositions, we expand the supertrace in \cref{eq:weq} up to the second order in $\widetilde{\Gamma}_{k, \p}$ (or in $\widetilde{\Gamma}_{k, p_0}$)
\begin{align}
\label{eq:flow_expansion}
\notag
  \STr{ \ln(\Gamma_k^{(2)}[\phi_{\p}]+R_k)} \approx  \STr{ \ln(\Gamma_k^{(2)}+R_k)}    &+   \STr\left\{{ (\Gamma_k^{(2)}+R_k)^{-1}  \widetilde{\Gamma}_{k,\p}^{(2)}}\right\} \\& - \frac{1}{2} \STr\left\{{(\Gamma_k^{(2)}+R_k)^{-1} \widetilde{\Gamma}_{k, \p}^{(2)}  (\Gamma_k^{(2)}+R_k)^{-1}  \widetilde{\Gamma}_{k,\p}^{(2) }}\right\},
\end{align}
where all $\Gamma^{(2)}$ matrices, in the absence of the fermionic background, take a block-diagonal form
\begin{align}
    A = \begin{pmatrix}
 [A]^{(\mathcal{B})} & 0 \\
 0 & [A]^{(\mathcal{F})} \\
\end{pmatrix}.
\end{align}
The supertrace operation is defined as
\begin{align}
\label{eq:str}
  \STr A =  \Tr [A]^{(\mathcal{B})} -  \Tr[A]^{(\mathcal{F})}.  
\end{align}

\subsection{Bosonic contribution}
The inhomogeneous part of the Hessian, obtained as the second functional derivative of $\Gamma_k$ with respect to the bosonic fluctuations, \cref{eq:nonlin}, and expanded to linear order in ${\mathtt a}, {\mathtt s}$, (sufficient for deriving the desired flows), is given by
\begin{align}
         [\widetilde{\Gamma}_{k,\p}^{(2)}(Q,Q')]^{(\mathcal{B})} =&  \beta (2\pi)^d \delta_{q_0+q_0', 0}   \big[ {\mathtt a} \, \delta(\q+\q' - \p)  +{\mathtt a}^* \, \delta(\q+\q' + \p) \big]  \begin{pmatrix}
 A_S & 0 \\
 0 & A_V \\
\end{pmatrix},
\end{align}
for the background  \cref{eq:p_vev}, and 
\begin{align}
\notag
  [\widetilde{\Gamma}_{k, p_0}^{(2)}(Q, Q')]^{(\mathcal{B})} =& \beta (2\pi)^d  \delta(\q +\q')   \big[ {\mathtt a} \, \delta_{q_0+q_0' - p_0, 0}  +{\mathtt a}^* \, \delta_{q_0+q_0' + p_0, 0}\big]  \begin{pmatrix}
 A_S & 0 \\
 0 & A_V \\
\end{pmatrix} \\
 & \hspace{10ex}+ (2\pi)^d \beta \delta(\q +\q')   \big[ {\mathtt s} \, \delta_{q_0+q_0' - p_0, 0}  +{\mathtt s}^* \, \delta_{q_0+q_0' + p_0, 0}\big]  \begin{pmatrix}
B_S & 0 \\
 0 &B_V \\
\end{pmatrix},
\end{align}
for the background  \cref{eq:p0_vev}. Here the block matrices are given by
\begin{align}
  A_S &=   \V'' \sqrt{2  \rho_1}  \begin{pmatrix}
 0 & 1 \\
 1 & 0 \\
\end{pmatrix},  \quad A_V =   \frac{2  \W}{N} \sqrt{2 \rho_1} \begin{pmatrix}
 0 & 1 \\
 1 & 0 \\
\end{pmatrix}  \otimes \id_{n_{\zeta}},
\end{align}
and 
\begin{align}
B_S &=   \V'' \sqrt{2  \rho_1}  \begin{pmatrix}
b_{\sigma}& 0 \\
 0& 1 \\
\end{pmatrix}, \quad
B_V =   \frac{2  \W}{N} \sqrt{2 \rho_1} \begin{pmatrix}
b_{\zeta} & 0 \\
 0 &b_{\pi} \\
\end{pmatrix}  \otimes \id_{n_{\zeta}},
\end{align}
with the diagonal elements 
\begin{equation}
\begin{split}
    \label{eq:b_coeff}
      b_{\pi} =  N \V''/2 \W, \quad 
    b_{\sigma} =  3 + 2 \rho_1 \V'''/\V'',  \quad 
      b_{\zeta} =  2+ 2\rho_1\W'/\W +b_{\pi}.  
\end{split}
\end{equation}
The flow equations thus  consist of two independent contributions: $S$-terms (corresponding to the ``Scalar'' modes $\alpha$ and $\sigma$ ) and $V$-terms (``Vector'' modes $\vec{\pi}$ and $\vec{\zeta}$ ), and are given by 
\begin{subequations}
\begin{align}
\dot{\Z}^{(\mathcal{B})}  &= \frac{1}{4 d}   \hat{\partial}_t \!  \! \sum_{\lambda = S,V}\int_{Q}\tr \Big(\frac{\partial}{\partial q_i} G_k^{(\lambda)}(Q) A_{\lambda} \frac{\partial}{\partial q_i} G_k^{(\lambda)}(Q) A_{\lambda} \Big), \\[1.5ex]
\dot{\A}^{(\mathcal{B})}  &= \frac{1}{2}   \hat{\partial}_t \!  \! \sum_{\lambda = S,V}\int_{Q}\tr \Big(  G_k^{(\lambda)}(Q) A_{\lambda} \frac{\partial}{\partial q_0} G_k^{(\lambda)}(Q) B_{\lambda} \Big). 
\end{align}
\end{subequations}
Applying the cutoff  \cref{eq:litimb},  we derive the bosonic contributions to the flow equations \cref{eq:der_coeff_flow_b}.

\subsection{Fermionic contribution}
The inhomogeneous additives to the fermionic part of the Hessian are given by 
\begin{align}
\label{eq:z1_pert}
         [\widetilde{\Gamma}_{k,\p}^{(2)}(Q,Q')]^{(\mathcal{F})} =& \frac{i g_k}{\sqrt{2N} } \beta (2\pi)^d \delta_{q_0, q_0'}   \big[ {\mathtt a} \, \delta(\q-\q' - \p)  +{\mathtt a}^* \, \delta(\q- \q' + \p) \big] \id_2 \otimes \idj, 
\end{align}
for the background  \cref{eq:p_vev}, and 
\begin{align}
\notag
  [\widetilde{\Gamma}_{k, p_0}^{(2)}(Q, Q')]^{(\mathcal{F})} =& \frac{g_k}{\sqrt{2N} }  \beta (2\pi)^d \delta(\q - \q')  \left\{ \big[ ({\mathtt s}^* - i {\mathtt a}^* )\, \delta_{q_0-q_0' - p_0, 0}  +({\mathtt s} - i {\mathtt a} )\, \delta_{q_0-q_0'+ p_0, 0}\big]  \begin{pmatrix}
 -\idj & 0 \\
 0 & 0 \\
\end{pmatrix}\right. \\
 & \hspace{10ex}+ \left. \big[ ({\mathtt s}^* + i {\mathtt a}^* )\, \delta_{q_0-q_0' - p_0, 0}  + ({\mathtt s} + i {\mathtt a} ) \, \delta_{q_0-q_0' + p_0, 0}\big]  \begin{pmatrix}
0 & 0 \\
 0 & \idj \\
\end{pmatrix}\right\},
\label{eq:z2_pert}
\end{align}
for the background \cref{eq:p0_vev}.
Performing the computation of \cref{eq:z1,eq:z2} using the expansion \cref{eq:flow_expansion} and the perturbative expressions \cref{eq:z1_pert,eq:z2_pert}, we derive the fermionic part of the flows for the coefficients of the derivative expansion
\begin{align}
 \dot{\Z}^{(\mathcal{F})}  & = \frac{g_k^2}{4d}  \,  \hat{\partial}_t  \int_{Q} \frac{1}{(q_0^2 + E_{\q, k}^2)^2} \left(\frac{\partial \xi_{\q, k} }{\partial q_i} \right)^2, \quad     \dot{\A}^{(\mathcal{F})}  = \frac{g_k^2}{2}  \,  \hat{\partial}_t  \int_{Q} \frac{\xi_{\q, k}}{(q_0^2 + E_{\q, k}^2)^2},   
\end{align}
where $ \xi_{\q, k} = \xi_{\q} + R^{(\mathcal{F})}_k$ and $ E_{\q, k}^2 = \xi_{\q, k}^2 + g_k^2 \rho_1/N$. Applying the cutoff \cref{eq:litimf}, we ultimately obtain the result \cref{eq:der_coeff_flow_f}. 

\end{widetext}

\section{\label{app:numeric}Numerical treatment}

A typical approach to solving partial differential equations numerically  is to perform a spatial discretization on an appropriate grid, followed by integration of the resulting system of ordinary differential equations (ODEs). This procedure, known as the method of lines, has been employed in our computations. Unfortunately, established techniques from numerical hydrodynamics \cite{Grossi23,Koenigstein1,Koenigstein2} are not directly applicable in this case, as the system of flow equations derived in our study cannot be cast in the form of conservation laws.

The potential varies significantly in the region where nontrivial extrema occur but changes gradually across the rest of the domain. To accurately resolve such regions, it is essential to ensure that a sufficient number of grid points lie within them. One way to achieve this is to remap the domain to a new coordinate system in which the solution varies more slowly. In this coordinate system, standard high-order finite-difference approximations can be applied to the first and second derivatives.

Let $x$ be a variable (in our case, $x = \Delta/\mu$) defined over the interval $[-x_0, x_0]$. Our goal is to construct a non-uniform mesh $\{x_i\}^{2 L-1}_{i=0}$ that more densely covers a target subinterval  $[-x^*, x^*] \subset [-x_0, x_0]$. In this work, we use the following mapping to a new variable $y$:
\begin{align}
\label{eq:map}
    x = x_0 \frac{\sinh(a y)}{\sinh a},
\end{align}
where $a$ is a scaling factor. The mesh points $y$ are uniformly distributed over the interval $[-1,1]$, forming a grid $\{y_i\}^{2L-1}_{i=0}$ with $y_i =  (i-L+1/2)  h$ and $ h = 1/(L-1/2)$.  We intentionally exclude the origin ($y=0$) from the mesh to avoid division by zero during numerical computations, although the original equations remain well-defined at this point. As $a \to 0$, the non-uniform grid smoothly transforms into the uniform one. As $a$ increases, the mesh points become more concentrated in a subdomain around the origin, the typical size of which is $x^* \sim 1/a$. The first and second derivatives are then rewritten in terms of  $y$. Derivatives of $y$ with respect to  $x$ are evaluated analytically for the mapping function \cref{eq:map}, while $y$-derivatives are approximated using the 7-order central differences (with accuracy $\mathcal{O}(h ^6)$), except  near the boundaries, where 3- and 5-order  central differences are used. The resulting system of ODEs is integrated from the UV scale $\Lambda$ down to an IR scale $k_0$ using an adaptive explicit Runge--Kutta method. Note that the system of ODEs is stiff, and it is not possible to reach arbitrarily small values of $k_0$, as this would require extremely small step sizes that may fall below machine precision. Therefore, it is essential to strike a balance between acceptable accuracy---determined by the grid size $L$, the size of the discretization domain $x_0$ and the IR scale $k_0$, below which the evolution of coupling constants is neglected---and the numerical stability of the integration algorithm.

The error estimates for the thermodynamic quantities presented in \cref{tab:thermo_quantities} are based solely on the accuracy of the discretization scheme. In this study, we do not account for potential variations in the numerical results arising from the choice of the ansatz \cref{eq:lpa} or the form of the regulator functions \cref{eq:litimb,eq:litimf}.  We varied the number of grid nodes $L \sim 100 - 300$ and the domain size $x_0 \sim 5-10$, ensuring that the resulting values of the presented parameters remain stable and exhibit only insignificant variations. 

Of course, a more suitable approach for stiff ODEs would be to use fully implicit integration schemes, which require pre-computation of the Jacobian. However, for systems as cumbersome as the one considered here, manually deriving and specifying the Jacobian is impractical. This highlights the potential utility of automatic differentiation techniques, which compute derivatives directly within the integration routine, as well as alternative strategies such as the Jacobian-free Newton–Krylov algorithm \cite{KNOLL2004357}. We plan to explore such approaches in future work.

\bibliography{paper}

\end{document}